\documentclass[12pt,aps,pra,preprint,superscriptaddress]{revtex4-1}

\usepackage{graphicx}% Include figure files
\usepackage[caption=false]{subfig}
\usepackage{dcolumn}% Align table columns on decimal point
\usepackage{bm}% bold math
\usepackage{float}
\usepackage{graphicx,wrapfig,lipsum}
\usepackage[english]{babel}
\usepackage{amsmath}
\DeclareMathOperator{\Tr}{Tr}
\newcommand{\be}{\begin{equation}}
\newcommand{\ee}{\end{equation}}
\newcommand{\bea}{\begin{eqnarray}}
\newcommand{\eea}{\end{eqnarray}}

\begin{document}
%\preprint{APS/123-QED}

%
% \draft command makes pacs numbers print
%\draft
% ****** Start of file apssamp.tex ******
%
%   This file is part of the APS files in the REVTeX 4 distribution.
%   Version 4.0 of REVTeX, August 2001
%

\title{Atom-Field Entanglement in Cavity QED: Nonlinearity and Saturation}

%>>>> The author is responsible for formatting the
%  author list and their institutions.  Use  \skiplinehalf
%  to separate author list from addresses and between each address.
%  The correspondence between each author and his/her address
%  can be indicated with a superscript in italics,
%  which is easily obtained with \supit{}.

\author{Robert Rogers}
\affiliation{Department of Physics, Miami University, Oxford OH 45056}
\affiliation{Department of Physics, University of Oregon, Eugene OR 97403}
\author{Nick Cummings}
\affiliation{Department of Physics, Miami University, Oxford OH 45056}
\affiliation{Integration Management and Operations Division, NASA
Kennedy Space Center, SR 405, Florida 32899 }
\author{Leno M. Pedrotti}
\affiliation{Department of Physics, University of Dayton, Dayton OH 45469}
\author{Perry Rice}
\affiliation{Department of Physics, Miami University, Oxford OH 45056}

\date{\today}% It is always \today, today,

%%%%%%%%%%%%%%%%%%%%%%%%%%%%%%%%%%%%%%%%%%%%%%%%%%%%%%%%%%%%%
\begin{abstract}
We investigate the degree of entanglement between an atom and a driven cavity mode in the presence of dissipation. 
Previous work has shown that in the limit of weak driving fields, the steady state entanglement is proportional to the square of the driving intensity.  
This quadratic dependence is due to the generation of entanglement by the creation of pairs of photons/excitations. In this work we investigate the entanglement between an atom and a cavity in the presence of multiple photons. 
Nonlinearity of the atomic response is needed to generate entanglement, but as that nonlinearity saturates the entanglement vanishes. We posit that this is due to spontaneous emission, 
which puts the atom in the ground state and the atom-field state into a direct product state. An intermediate value of the driving field, near the field that saturates the atomic response, optimizes the atom-field entanglement.
In a parameter regime for which multiphoton resonances occur, we find that entanglement recurs at those resonances. In this regime, we find that the entanglement decreases with increaing photon number. 
We also investigate, in the bimodal regime, the entanglement as a function of atom and/or cavity detuning.  Here we find that there is evidence of a phase transition in the entanglement, 
which occurs at $2\epsilon/g \geq 1$.
\end{abstract}

\maketitle
%%%%%%%%%%%%%%%%%%%%%%%%%%%%%%%%%%%%%%%%%%%%%%%%%%%%%%%%%%%%%
\section{Introduction}
Over the last two decades studies into the foundations of quantum mechanics
have shown that entanglement is an inherently quantum mechanical property
that, among other things, is useful for description of prime product
factorization, search protocols, and quantum teleportation \cite{Nielsen-2000-book,JaegerBook}.
The measure of entanglement is well defined in the case of two interacting
qubits. Consider a two qubit state
\begin{equation}
  | \psi \rangle = C_{00} |00 \rangle + C_{01} |01 \rangle + C_{10} |10
  \rangle + C_{11} |11 \rangle
\end{equation}
If a system is in the product state
\begin{equation}
  | \psi \rangle = (A_0 |0 \rangle + A_1 |1 \rangle) (B_0 |0 \rangle + B_1
  |1 \rangle)
\end{equation}
then the probability amplitudes will satisfy
\begin{equation}
  \mathcal{E} \equiv C_{00} C_{11} - C_{01} C_{10} = 0
\end{equation}
If this relation is {\itshape{not}} satisfied then the state is entangled. The
concurrence $\mathcal{C} = \sqrt{2 \mathcal{E}}$ is a measure of entanglement
for two qubits \cite{Hill-1997-5022,Wooters-98-Ent}. The log negativity $\mathcal{E}_N$ is another
measure of entanglement that presents an upper bound to the entanglement of
distillation and is conveniently additive \cite{Vidal-02-LogNeg}. In the simple case of two qubits
$\mathcal{E}_N = \log_2  (|| \rho^{T_F} ||) = \log_2 (1 + \sqrt{2} 
\mathcal{E)}$.

Three qubits can be entangled in several ways. The qubits can be in a
Greenburger-Horne-Zeilinger \cite{GHZ90} state, a W state \cite{Durr-00-3qubitent}, or the entanglement can be
solely between various bipartite splits of the system (i.e. particle 1
entangled with 2 and 3 etc. ) \cite{Coffman-2000-052306}. A system of two coupled qubits that
interact with an environment via pumping and/or dissipation is similar to a
three-qubit system. In this case, the system is essentially tri-partite: qubit
1, qubit 2, and the environment. If we choose to trace or average over the
environmental degrees of freedom, the two qubits will be described by a mixed
state instead of a pure state.

There are many ways to measure entanglement \cite{Hordeckis-09-Ent-RMP}, many of which
require optimization of a convex function. In this paper we adopt a pragmatic
approach, introduced by Nha and Carmichael \cite{Nha-2004-120408}, that utilizes quantum
trajectory theory. Open quantum systems can be described by a reduced density
matrix that is obtained by tracing over environmental degrees of freedom.
Recall that in quantum trajectory theory, the density matrix takes the form

\begin{equation}
  \dot{\rho} = \mathcal{L} \rho = (\mathcal{L - S}) \rho + \mathcal{S} \rho
\end{equation}
where $\mathcal{L}$ is the Liouvillian operator defined by the master
equation, and $\mathcal{S}$ is any combination of system operators. For a
given choice of $\mathcal{S}$, the $\mathcal{L - S}$ part of the evolution can
be described by an effective Schroedinger equation, with a non-Hermitian
Hamiltonian evolving the conditioned state $| \psi_c \rangle$. This is
referred to as a quantum trajectory \cite{Carmichael-1993-book,Molmer-1996-49,Tian-92}. This evolution is punctuated by the
application of the $\mathcal{S}$ operator, which we refer to as a jump, at
times randomly chosen from a distribution that is determined by the current
state of the system. At every time step the trajectory must be normalized, as
a jump or non-Hermitian evolution results in nonunitary evolution. To
fully recover the statistical information about the system, one then averages
over a set of trajectories. Obviously, a different choice for $\mathcal{S}$
leads to a different unraveling and a different set of trajectories. One
common choice for $\mathcal{S}$ is that of direct detection, where, for
example, the spontaneous emission of an atom is monitored via a $4 \pi$
detector with perfect efficiency. This yields quantum trajectories conditioned
on this measurement record. Another common choice is that of homodyne
detection, where the number of jumps is very high over the characteristic
dissipatiive/driving rates of the system and one coarse grains the resulting
equation, resulting in a nonlinear Schroedinger equation for evolution.

Good reviews of this topic and its relationship to measurement theory can be
found in the works of Wiseman \cite{Wiseman-96-QT} and Jacobs and
Steck \cite{Jacobs-06-QTreview}. Nha and Carmichael proposed applying pure state
entanglement measures to the quantum trajectories, yielding a functional
definition of entanglement for open systems. They explicitly demonstrated that
the "amount" of entanglement may be {\itshape{different}} for different
choices of $\mathcal{S}$ by examining the results given by homodyne and direct
detection applied to their optical system. For weak fields the unraveling is
irrelevant, as the system is well approximated by a pure state for many
lifetimes, as the jump rate $P (t) dt = \gamma \langle \sigma_z (t) + 1 / 2
\rangle dt$ and $\sigma_z (t) \approx - 1$ for weak fields, so jump events are
relatively rare. It has been shown that while entanglement is different for different unravelings, for a given unraveling all measures of entanglement show similar behavior as system parameters are changed \cite{Guevara-90-012338}.

In this paper we use this method of examining entanglement in open quantum
systems and choose jump operators corresponding to direct detection. We take
the point of view that a quantum trajectory based on direct detection
monitoring could aid in understanding the amount of useful entanglement in a
quantum information processing protocol that utilizes direct detection for
state preparation and processing.

In Section II we describe the system to be studied. In Section III, results for
weak driving fields are presented and the validity of the weak driving-field
approximation is examined. The behavior of the log negativity for a system
driven on resonance with a field of arbitrary strength is treated in Section
IV. Results for the behavior of the log negativity for a system driven off resonance by
fields of a variety of strengths are presented in Sections V and VI.

\section{System Description}
 We use a python toolkit for modeling the dynamics of open quantum systems \cite{Johansson-2012-1760,Johansson-2013-1234}. The system we investigate is a single two-level atom located
at an anti-node of a field mode in an optical cavity, as shown in Fig. 1. The
cavity is driven by a photon flux $\epsilon$, one end mirror is "leaky" with a
field dissipation rate of $\kappa$, and the atom spontaneously emits out the
side of the cavity at the cavity-modified rate $\gamma$. The coupling between
the field mode and the atom is given by $g$.
\begin{figure} [!htb]
   \begin{center}
   \begin{tabular}{c}
   \includegraphics[height=6cm]{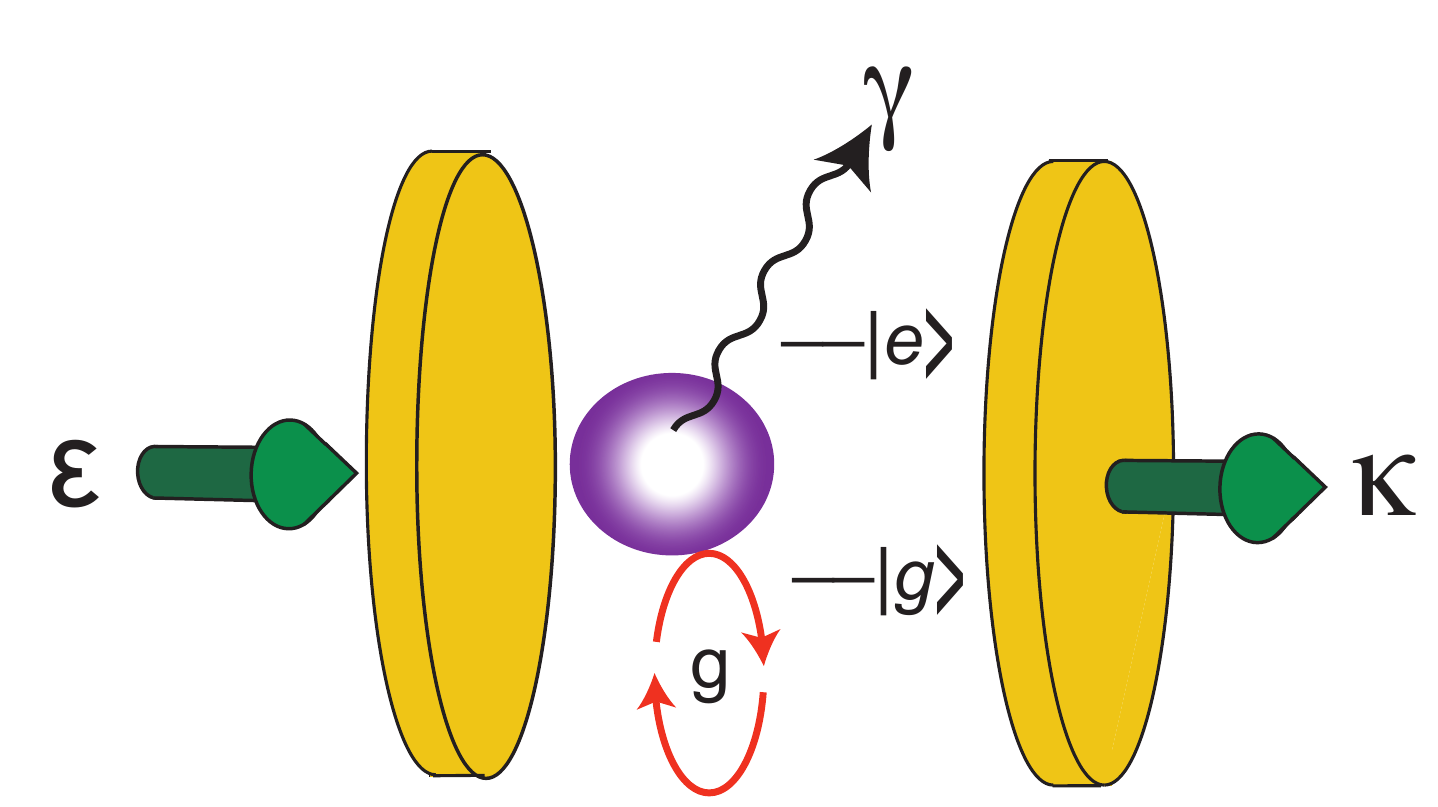}
   \end{tabular}
   \end{center}
   \caption[example]
%>>>> use \label inside caption to get Fig. number with \ref{}
%   { \label{fig:example}
{Single atom in a driven optical cavity with dipole coupling $g$, cavity field loss rate $\kappa$, and atomic spontaneous emission rate $\gamma$.}
   \end{figure}

The quantum trajectory wave function and Hamiltonian that characterize the
system are:
\begin{eqnarray}
  | \psi_c (t) \rangle & = & \left. \sum_n^{\infty} ( C_{g, n} (t) e^{- iE_{g,
  n} t} |g, n \rangle  + C_{e, n} (t) e^{- iE_{e, n} t} |e, n \rangle) \right.
  \label{psi}\\
  H & = & \frac{\hbar}{2} \Delta \sigma_z + \hbar \theta a^{\dagger} a + \hbar
  g \hspace{0.27em} (a^{\dagger} \sigma_- + a \sigma_+) - i \hbar \kappa
  a^{\dagger} a - i \hbar \frac{\gamma}{2} \sigma_+ \sigma_- + i \hbar
  \epsilon (a^{\dagger} - a) 
\end{eqnarray}
with collapse operators
\begin{eqnarray}
  \mathcal{A} & = & \sqrt{\kappa} a \\
  \mathcal{S} & = & \sqrt{\frac{\gamma}{2}} \sigma_- . 
\end{eqnarray}
associated with detection of photons exiting the output mirror and out the
side of the cavity, respectively. The indices $e$ and $g$ indicate the atom in
the excited or ground state, while $n$ is the number of photons in the mode.
The creation and annihilation operators for the field are $a^{\dagger}$ and
$a$, and the Pauli raising and lowering operators for the atom are $\sigma_+$
and $\sigma_-$. The detunings are given by $\Delta = 2 (\omega_{atom} -
\omega_{cav}) / \gamma$ and $\theta = (\omega_{drive} - \omega_{cav}) /
\kappa$.

\section{Weak Driving Field}

In the weak driving-field limit, the system reaches a steady-state wave
function of the form
\begin{equation}
  | \psi_{WF} \rangle = |0g \rangle + A_{1, g} |1g \rangle + A_{0, e} |0e
  \rangle + A_{2, g} |2g \rangle + A_{1, e} |1e \rangle \label{wavefunction}
\end{equation}
where the $A_{i, j}$ are known \cite{Carmichael-1991-73,Brecha-1999-2392}.

\

To lowest order in the weak driving-field approximation, $A_{1, g}$ and
$A_{0, g}$ are proportional to the driving field strength $\epsilon$. A
quantity that is key to understanding entanglement in the weak field limit is

\be
\frac{A_{1, e}}{A_{1, g} A_{0, e}}
  = q =\frac{(1+2C_1)}{(1+2C_1-2C^{'}_1)}\label{q}
\ee
with $C_1 = g^2 / \kappa \gamma$ and $C_1' = 2 C_1 \kappa / (2 \kappa +
\gamma)$. As noted above, the one-excitation amplitudes $A_{1, g}$ and $A_{0,
e}$ are proportional to the driving field $\epsilon$ and the two-excitation
amplitudes, $A_{2, g}$ and $A_{1, e}$, are proportional to the square of the
driving field, $\epsilon^2$. \cite{Carmichael-1991-73,Brecha-1999-2392}.

The key quantity $q$ is related to the cross correlation
\begin{equation}
  g^{(2)}_{TF} (0) = \frac{\langle a^{\dagger} a \sigma_+ \sigma_-
  \rangle}{\langle a^{\dagger} a \rangle  \langle \sigma_+ \sigma_- \rangle} =
  q,
\end{equation}
which is the probability of simultaneous detection of transmitted and
fluorescence photons scaled by the product of the probability of two
independent detections in transmission and fluorescence.

\

\

In the weak-field limit, $\epsilon / \kappa \ll 1$, the concurrence for this
system is found to be
\begin{equation}
  \mathcal{C} = - \left( \frac{\epsilon}{\kappa} \right)^4 \log_2 \left[
  \left( \frac{\epsilon}{\kappa} \right)^4 \right] \xi^2,
\end{equation}
where
\begin{equation}
  \xi = \frac{2 g}{\gamma (1 + 2 C_1)^2}  (q - 1) .
\end{equation}
These relations hold provided $(\epsilon / \kappa)^2 \ll | \xi | < 1$. The log
negativity is found simply from the concurrence.

\

We now investigate the range of validity of the weak-field approximations
given above. Figure 2 shows the fidelity $\mathcal{F}$ defined as
\begin{equation}
  \mathcal{F} = \Tr (1 - \rho_c^2)
\end{equation}
We have performed an average over many trajectories to obtain an
ensemble average. We see that a pure state approximation is valid for driving
fields well below the strength of the saturation field. We also plot the norm
of the distance between the state we calculate using trajectories for an
arbitrary field strength and the pure state obtained in the weak field limit.
This norm $\mathcal{D}$ has the form
\begin{equation}
  \mathcal{D} = | \langle \psi_{WF} | \psi_c \rangle |^2
\end{equation}

Again we find good agreement for field strengths up to half the saturation
field. That this agreement is sensible can be understood as follows. The large
component of the state corresponds to the vacuum state of the field and the
ground state of the atom. Hence, at field strengths below half the saturation
field, the mean photon number is very small (close to zero), and the atomic
inversion is very close to $- 1$, i.e. the atom in the ground state. Hence,
the purity of the state is not unexpected for low field strengths, though the
similarity between the actual state and the weak field state approximation is,
perhaps, a bit surprising. However, we posit that $| \psi_c \rangle = |0
\rangle + d| \text{1-excitation} \rangle$ and $| \psi_{WF} \rangle = |0
\rangle + d' | \text{1-excitation} \rangle$ giving a distance that is
approximately $|d - d' |$, which is proportional to $\epsilon^2$ and so is
small compared to $1$ for weak driving fields. Also, as the driving field
strength becomes larger, the state of the system is no longer in a
quasi-steady state for many lifetimes as it would be in the weak driving
field approximation and the rate of spontaneous emission jumps is on the order
of the lifetime. The spontaneous emission rate is greatest at saturation and
each spontaneous emission event destroys entanglement.

\begin{figure} [!htb]
   \begin{center}
   \begin{tabular}{c}
   \includegraphics[height=12cm]{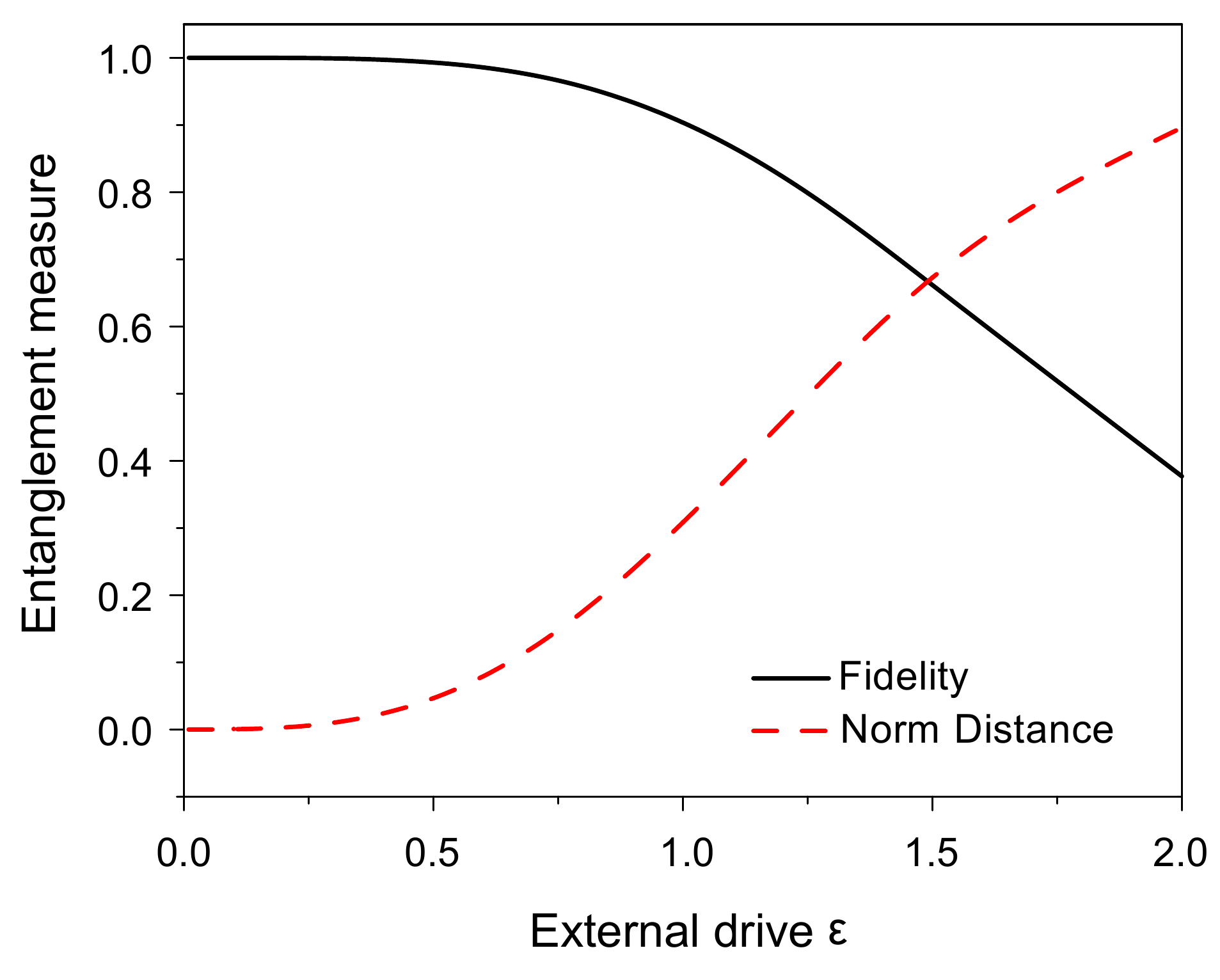}
   \end{tabular}
   \end{center}
   \caption[example]
%>>>> use \label inside caption to get Fig. number with \ref{}
%   { \label{fig:example}
{Comparison of the wavefunction to the one obtained analytically in the week field limit via two measures, the fidelity and norm of the distance, both defined in the text.}
   \end{figure}

\section{Log Negativity on Resonance}

We now consider entanglement in the presence of driving fields of arbitrary
strength that are resonant with both the atom and the cavity. In Fig. 3, we
plot the population inversion of the atom and intracavity photon number as a
function of resonant driving field, with coupling and dissipation rates of
similar magnitude, and with the driving field scaled so that when $\epsilon =
1$, the intensity in the cavity is equal to the saturation intensity of the
atom. We see typical behavior, the photon number rising as the driving field
is increased and becoming linear for high fields after the atom saturates
\cite{Bonifacio-78-OB-BL}. Low and high fields in this case being $\epsilon \ll 1$ and
$\epsilon \gg 1$ respectively. The atomic inversion increases with increasing
driving field and pins at the value of zero for high fields.

\begin{figure} [!htb]
   \begin{center}
   \begin{tabular}{c}
   \includegraphics[height=12cm]{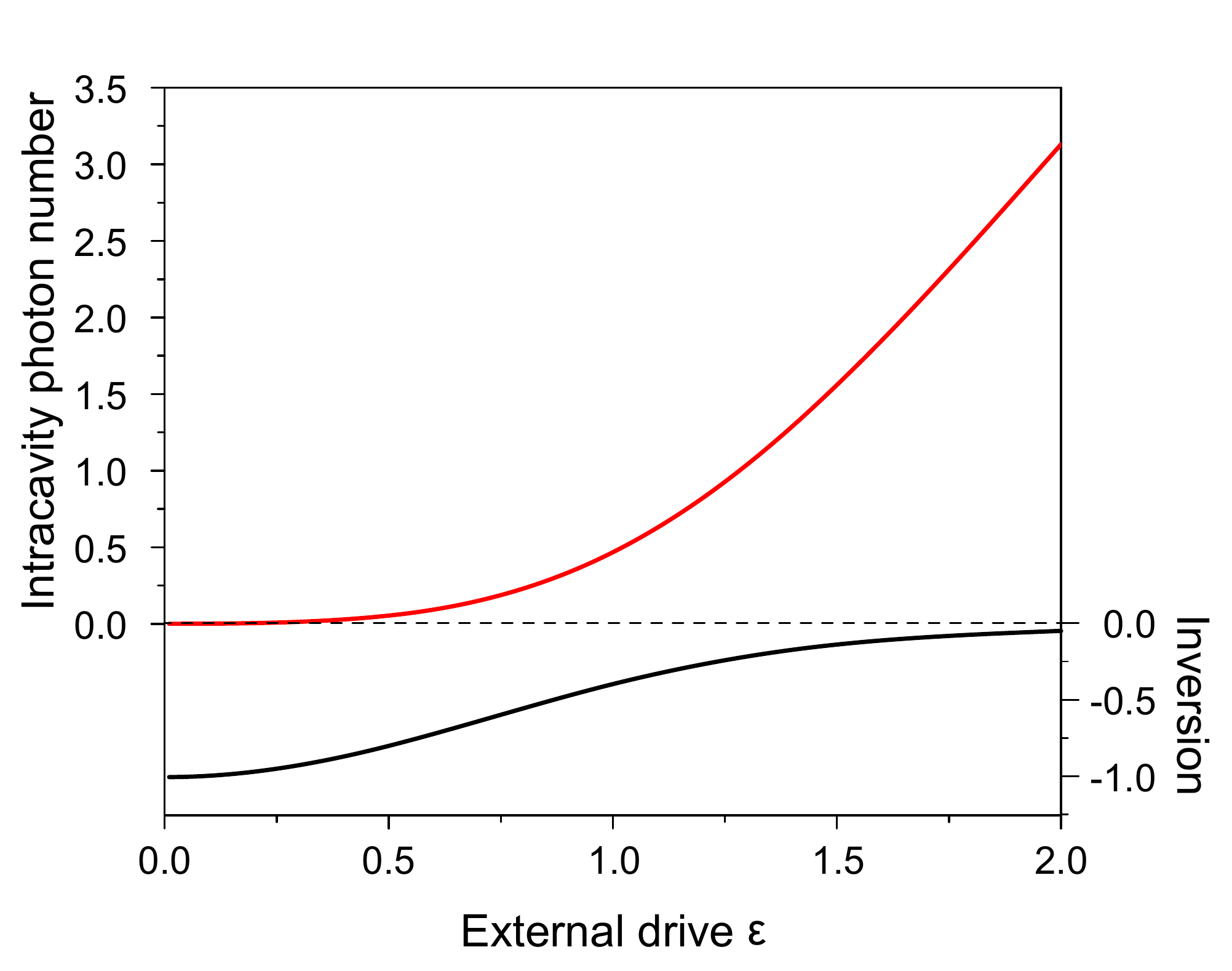}
   \end{tabular}
   \end{center}
   \caption[example]
%>>>> use \label inside caption to get Fig. number with \ref{}
%   { \label{fig:example}
{Average intracavity photon number and atomic population inversion as a function of driving field for $\kappa=\gamma=g=1.0$.}
   \end{figure}

We use as our measure of entanglement the logarithmic negativity defined as
\begin{equation}
  \mathcal{E}_N = \log_2 (\| \rho^{T_F} \|) .
\end{equation}
Here $\| \rho^{T_F} \|$ is the trace norm of the partial transpose of the
density matrix with respect to the field, which is formed in the following
manner. The wavefunction from our trajectory theory is used to calculate a
density matrix for the atom-field system. This is then used in constructing
$\rho^{T_F}$. We truncate the basis for the Fock states of the field mode at a
number well above the average photon number as predicted by semiclassical
theories, and then adjust that upper truncation level to assure that our
results are insensitive to that truncation. We obtain similar results using
the negativity $\mathcal{N} = (1 / 2)  (|| \rho^{T_F} || - 1)$ and concurrence
$\mathcal{C}$; we use the log negativity for ease of calculation.

Figure 4(a) shows the temporal development of the log negativity as a function
of time for a variety of driving field strengths. Here we see a well defined
steady-state entanglement value, with a preceding maximal value. There are
also oscillations that are roughly at the Rabi frequency for larger driving
fields, and which are written into the dynamics via the eigenvalues of the time
evolution operator. In Figure 4(b), we plot the log negativity as a function of
driving strength for several times. We see that at long enough times, there is
indeed an optimal driving field near $\epsilon = 1$.
 
\begin{figure}[!htb]
  \begin{center}
 
     \subfloat[]{%
       \includegraphics[width=0.47\textwidth]{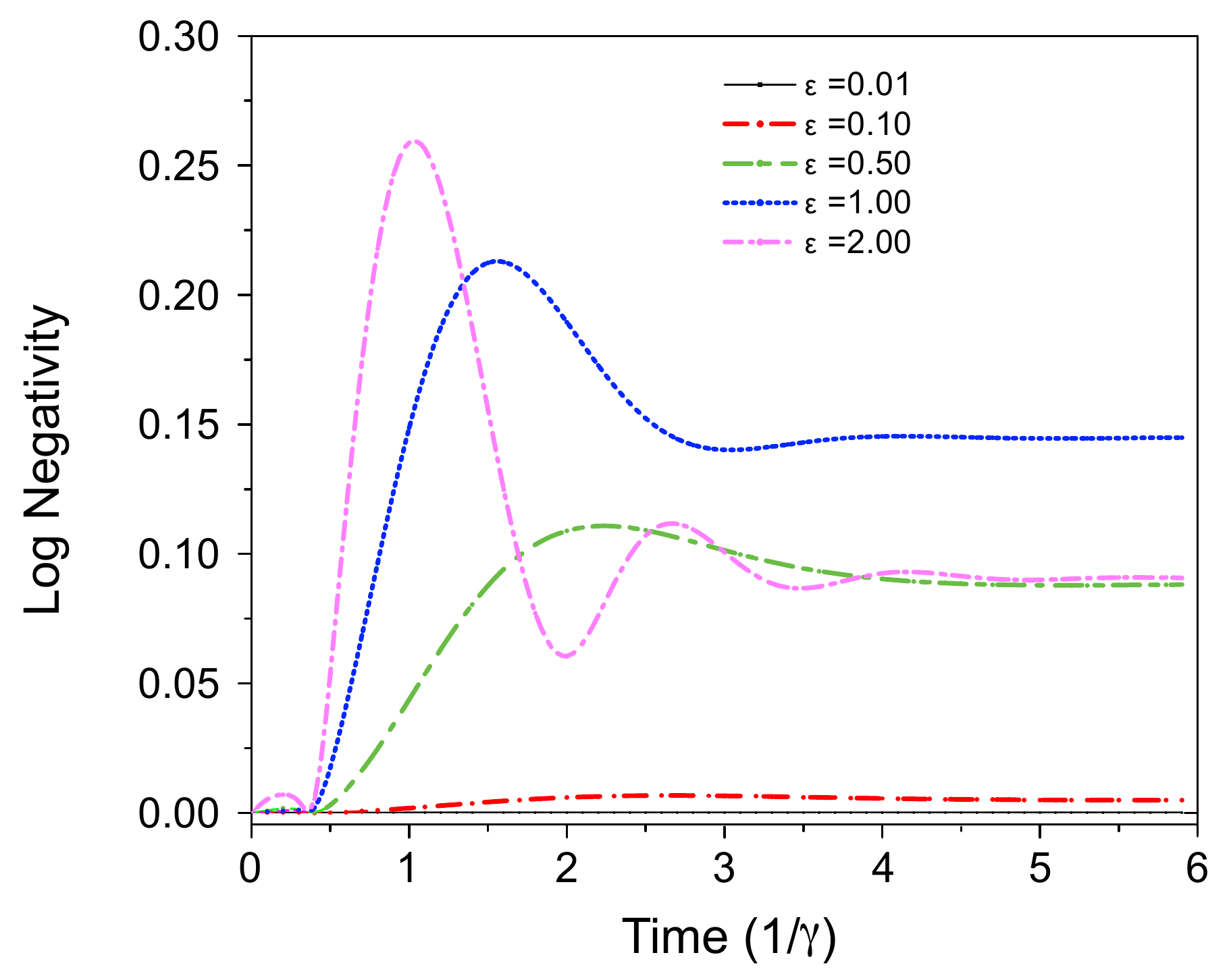}
     }
     \hfill
     \subfloat[]{%
       \includegraphics[width=0.47\textwidth]{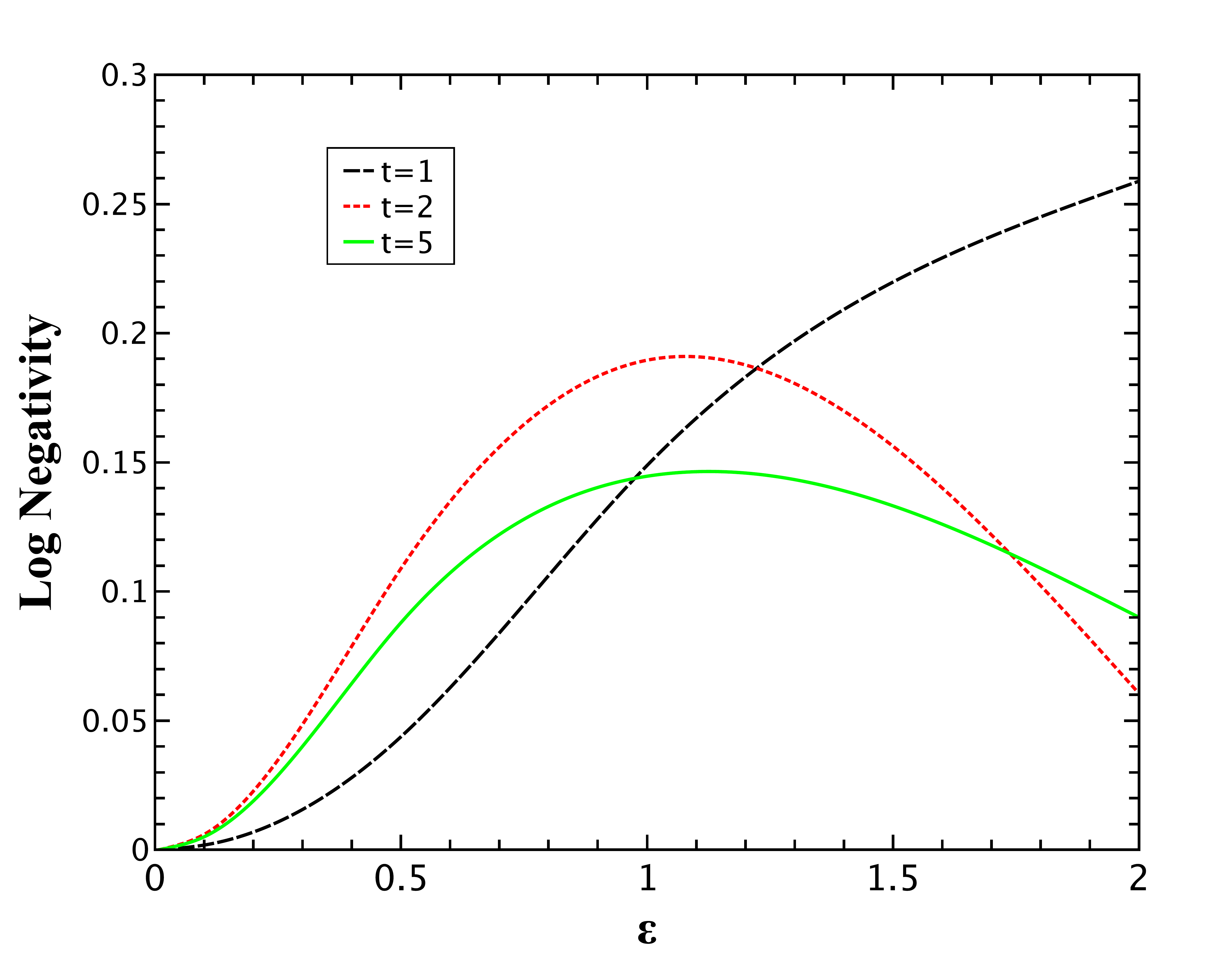}
     }
     \caption{(a) Log negativity as a function of time for various driving field strengths. (b) Log negativity as a function of driving field strength for $\gamma t=1.0$, $2.0$, and $5.0$.}
  \end{center}  
     \end{figure}

   Figure 5 shows the log negativity as a function of the driving field strength.
We plot both the steady state value, reached several atomic/cavity lifetimes
after the system is started in the ground state, as well as the maximum value
obtained over that same time period. We find that the optimum value for the
driving field strength is equal to the field that would saturate the atom,
$\epsilon \approx 1.0$. Beyond that field strength the entanglement decreases,
ultimately to zero.
\begin{figure} [!htbp]
   \begin{center}
   \begin{tabular}{c}
   \includegraphics[height=10cm]{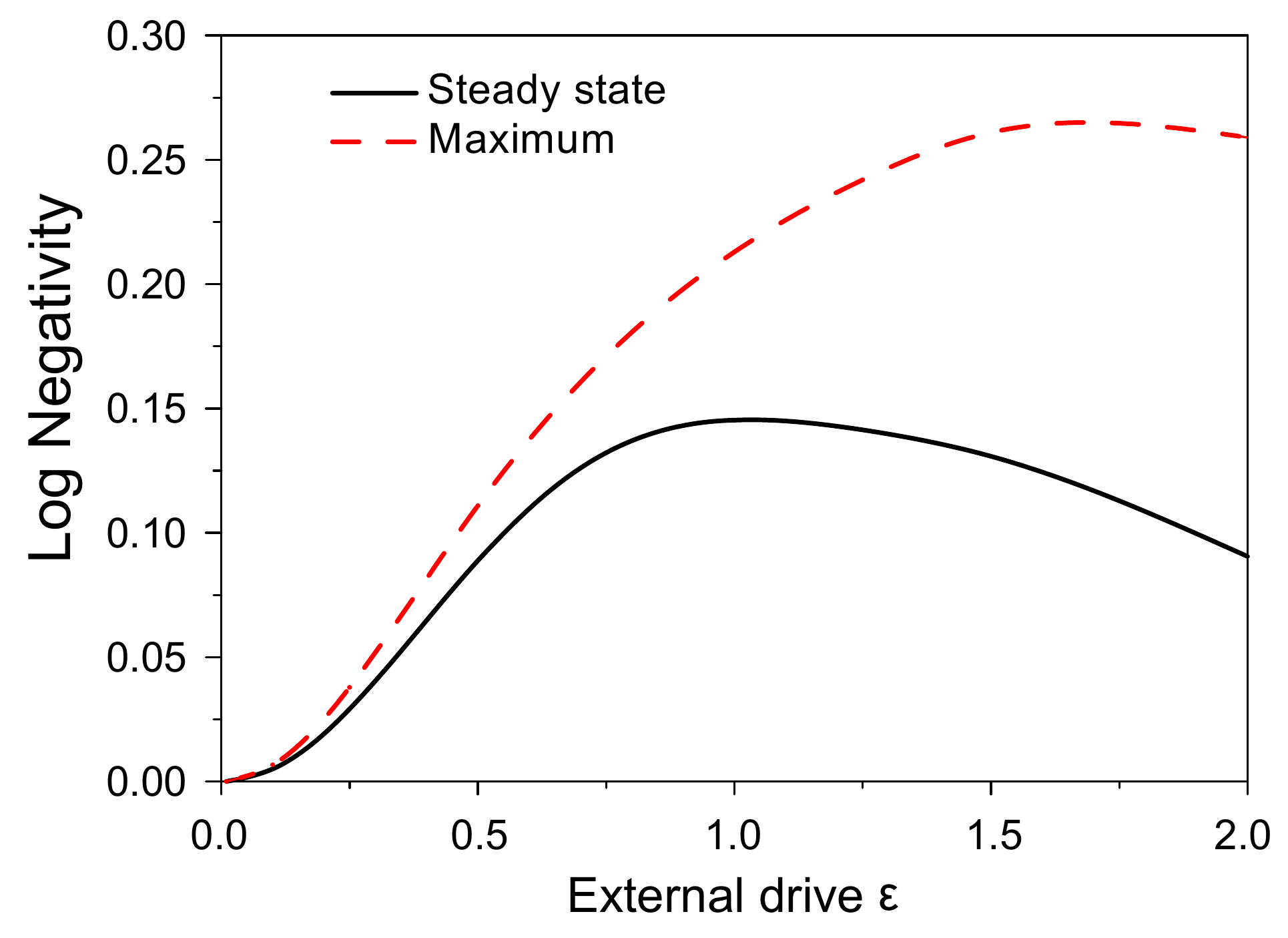}
   \end{tabular}
   \end{center}
   \caption[example]
%>>>> use \label inside caption to get Fig. number with \ref{}
%   { \label{fig:example}
{Maximum and steady-state value of the log negativity as a function of field strength.}
   \end{figure}

As two oscillators coupled together with a linear coupling will not become
entangled, we know that the nonlinearity of the atom is crucial to the
formation of atom-field entanglement. However, when that nonlinearity
saturates, the entanglement decreases. If one replaces the atom with a harmonic
oscillator, it is driven to a coherent state and $q = 1$. Essentially the atom
is in a mixed state with equal probabilities of being in the excited and ground
states, and the atom-field state factorizes. The physical mechanism responsible
for the loss of entanglement is spontaneous emission, which is maximized for a
saturated atom.
   
\section{Log Negativity Off Resonance}

The entanglement off resonance and for relatively high driving
fields is examined in this section. In Fig. 6 we see that the general behavior in which the log negativity vanishes as the atom
saturates holds over different field-cavity detunings $\theta$. However, when
$\theta \neq 0$ there is an asymmetry, and the log negativity persists at
slightly higher driving fields.

\begin{figure}[!htbp]
  \begin{center}
 
     \subfloat[]{%
       \includegraphics[width=0.47\textwidth]{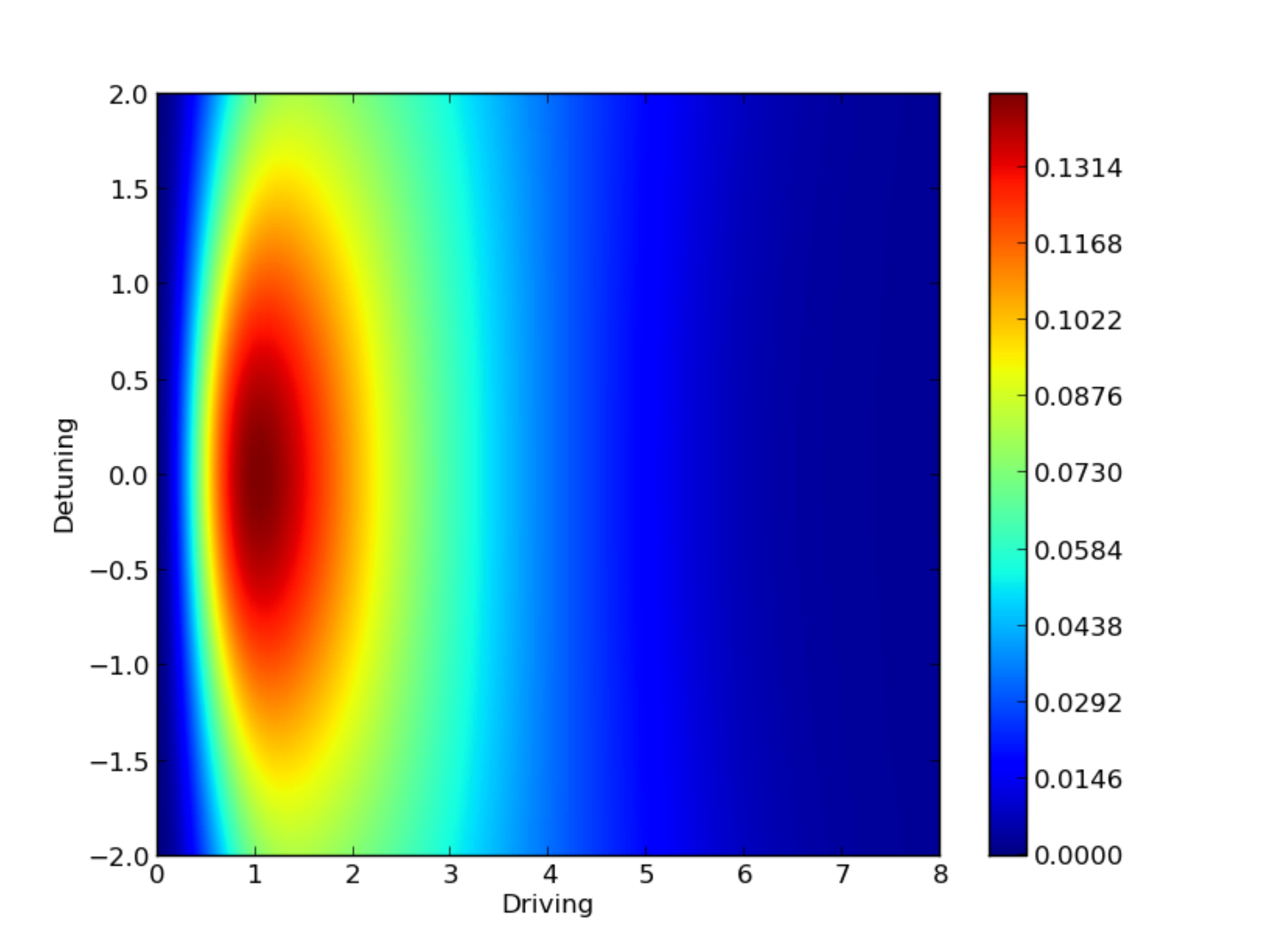}
     }
     \hfill
     \subfloat[]{%
       \includegraphics[width=0.47\textwidth]{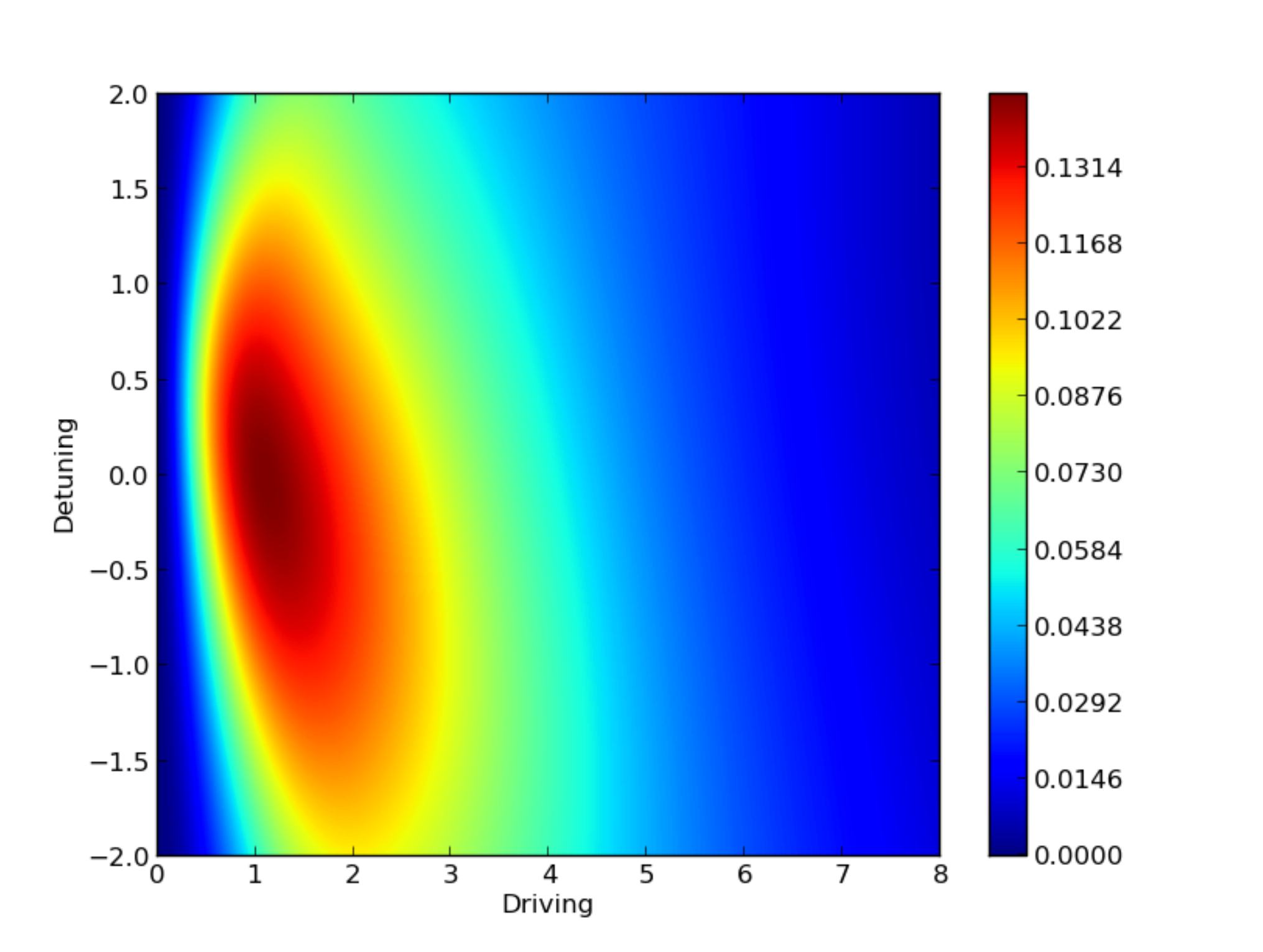}
     }
     \caption{The log-negativity with (a) $\theta = 0$ and (b) $\theta = 0.5$ . Parameters used are $\kappa,\gamma$ and $g = 1$. 100 states of the cavity are kept.}
  \end{center}  
\end{figure}

\begin{figure}[!htbp]
  \begin{center}
 
     \subfloat[]{%
       \includegraphics[width=0.47\textwidth]{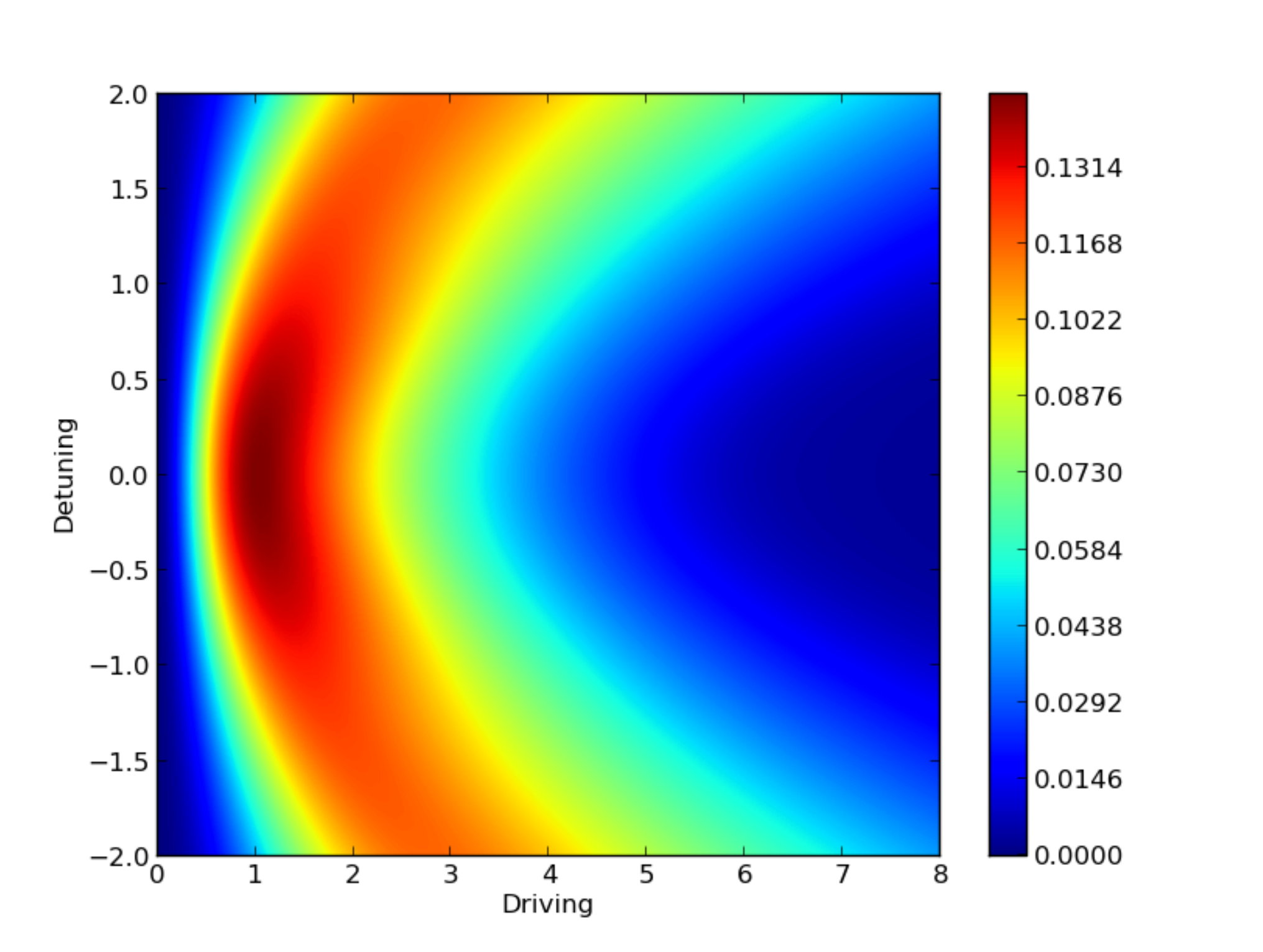}
     }
     \hfill
     \subfloat[]{%
       \includegraphics[width=0.47\textwidth]{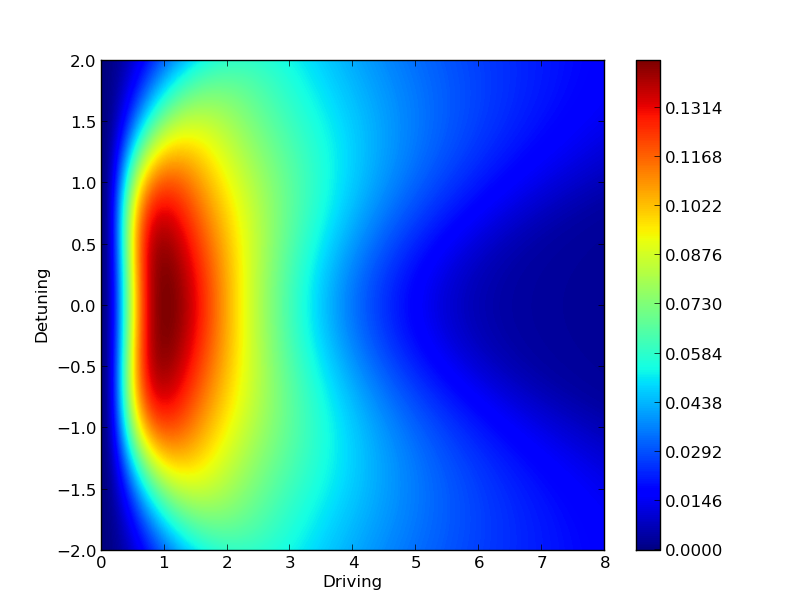}
     }
     \caption{The log-negativity with (a) $\theta = -\Delta$ and (b) $\theta = \Delta$. Parameters used are $\kappa,\gamma$ and $g = 1$.}
  \end{center}  
\end{figure}

Figure 7 shows that when $\theta = \pm \Delta$ this persistence is either
enhanced or suppressed. Explanation for this behavior can be found in the
examination of the effect $\Delta$ and $\theta$ have on the vacuum Rabi
frequency. Following Brecha, Rice, and Xiao \cite{Brecha-1999-2392} we note that
including these detunings effectively makes $\kappa$ and $\gamma$ complex,
\begin{eqnarray}
  \kappa \to \kappa (1 + i \theta) = \tilde{\kappa} \\
  \gamma \to \gamma (1 + i \Delta) = \tilde{\gamma} . 
\end{eqnarray}
The vacuum Rabi frequency becomes
\begin{eqnarray}
  \tilde{\Omega} = \sqrt{\left( \frac{1}{4} \right)  \left( \tilde{\kappa} -
  \frac{\tilde{\gamma}}{2} \right)^2 - g^2} 
\end{eqnarray}
Note that the relative signs of $\kappa$ and $\gamma$ matter a great deal. If
$\theta = - \Delta$ the terms enhance one another, and hence we observe the
behavior in Fig. 7(a) while, if the $\theta = \Delta$, we see behavior similar 
to driving on resonance, producing the results in Fig. 7(b). The behavior in Fig. 7(a) is due to the persistence of vacuum Rabi oscillations when the detunings "cancel" in this manner. One effectively tunes to a vacuum Rabi sideband, and sees two-level behavior as in Tian et. al.
\cite{Tian-92}.

\section{Strong Coupling and Driving Off Resonance}

In this section the behavior of the system for extremely high coupling between the
cavity and atom is examined. Though the results presented
above seem to imply that entanglement vanishes for fields much larger than
the field that saturates the atom, we instead find that, when coupling is of
sufficiently high magnitude, there is a regime where the entanglement
reappears. This behavior in the average photon number has been noted before,
and is due to the $g \sqrt{n}$ Rabi splitting via strong coupling and/or high
driving fields. In general, these conditions give rise to multi-photon
processes {\cite{Shamailov2010766}. As various multi-photon resonances are
obtained, there is a regime in which the system effectively behaves like
a two-level system, resulting in entanglement as in the resonant weak driving-field
case. The results of our calculations are shown in Fig. 8.

\begin{figure}[!htbp]
  \begin{center}
 
     \subfloat[]{%
       \includegraphics[width=0.47\textwidth]{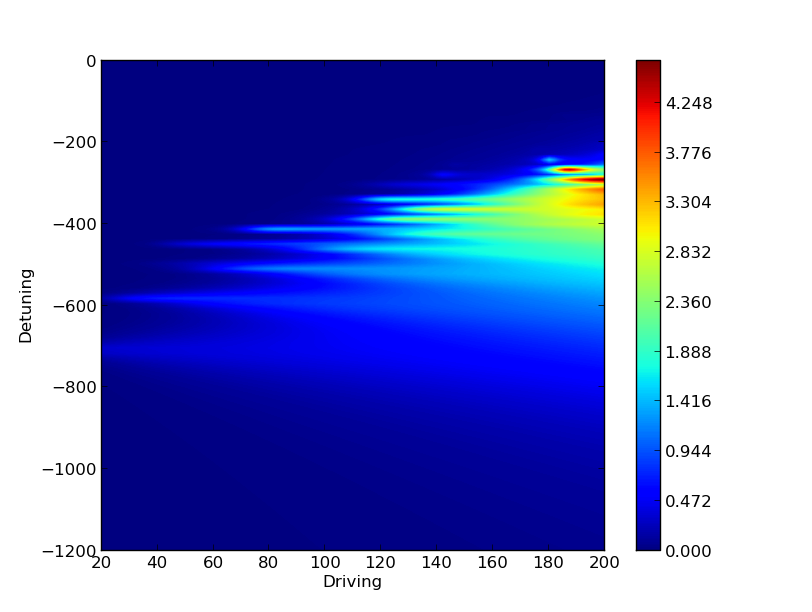}
     }
     \hfill
     \subfloat[]{%
       \includegraphics[width=0.47\textwidth]{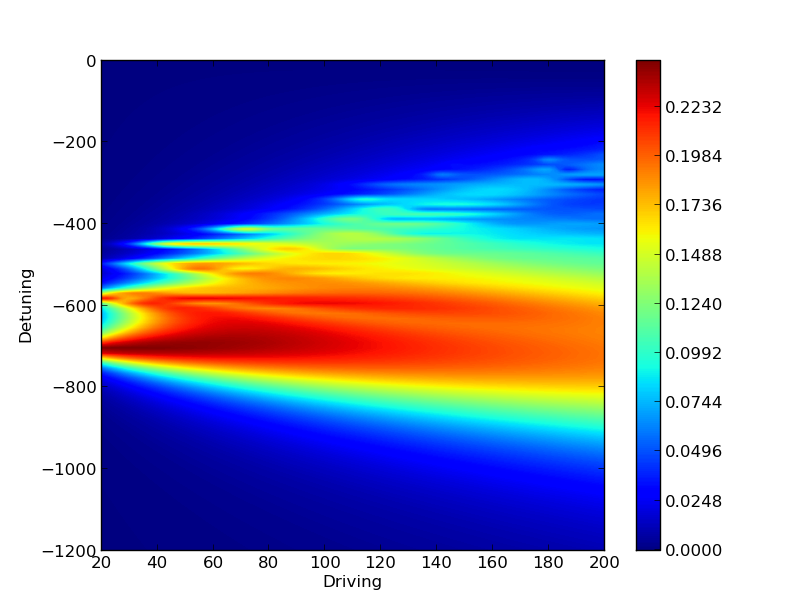}
     }
     \caption{Intracavity photon number (a) and log negativity (b) for high driving fields with $\theta = \Delta$. Parameters used are $\kappa = 1$, $\gamma=2$, and $g = 1000$. 15 states of the cavity kept.}
  \end{center}  
\end{figure}

Notice that for larger photon number, the accompanying entanglement is
smaller. This is due to the increase of the excited state population of the
effective two level system with increasing intracavity photon number. Again,
there is less entanglement if the atom saturates.

Another regime of interest is near the phase transition that occurs where $2
\epsilon / g = 1$ \cite{Alsing-91-DressedState,Alsing-92-DSS,Carm-15-X}. The entanglement in this regime has
been studied by Gea-Banacloche et. al. {\cite{Gea-Banacloche-2005-053603} outside the secular
approximation used by Alsing et. al. \cite{Alsing-92-DSS,Alsing-91-DressedState}, where one ignores the
difference between oscillations at $\sqrt{(\langle n \rangle \pm \langle n
\rangle^{1 / 2})} g$. We do not make the secular approximation, and we find
that the atom and field are entangled in the regime where $2 \epsilon / g >
1$, with a sharp demarcation between this regime and the regime at fields
below the phase transition. As for weak fields, when one drives the system
to saturation, spontaneous emission kills the entanglement. This behavior is
depicted in Fig. 9. One anticipates the presence of entanglement noting that Gea-Banacloche et. al. \cite{Gea-Banacloche-2005-053603} found that while no steady state exists, for individual trajectories one finds
\be
|\psi\rangle_{ss}=\frac{1}{\sqrt{2}}\left[|\psi_+^0(|\alpha_{ss}|,-\Phi_{ss})\rangle+e^{-i\theta}|\psi_-^0(|\alpha_{ss}|,\Phi_{ss})\rangle\right]
\ee
with
\be
|\psi_{\pm}^0 (r,\theta_0)\rangle=\frac{1}{\sqrt{2}}\left[e^{-i\theta_0}|e\rangle\pm|g\rangle\right]|re^{-i\theta_0}\rangle
\ee
which is clearly an entangled state. These persist until disturbed by a spontaneous emission event that disentangles the state.

\begin{figure}[H]
   \begin{center}
   \begin{tabular}{c}
   \includegraphics[height=6cm]{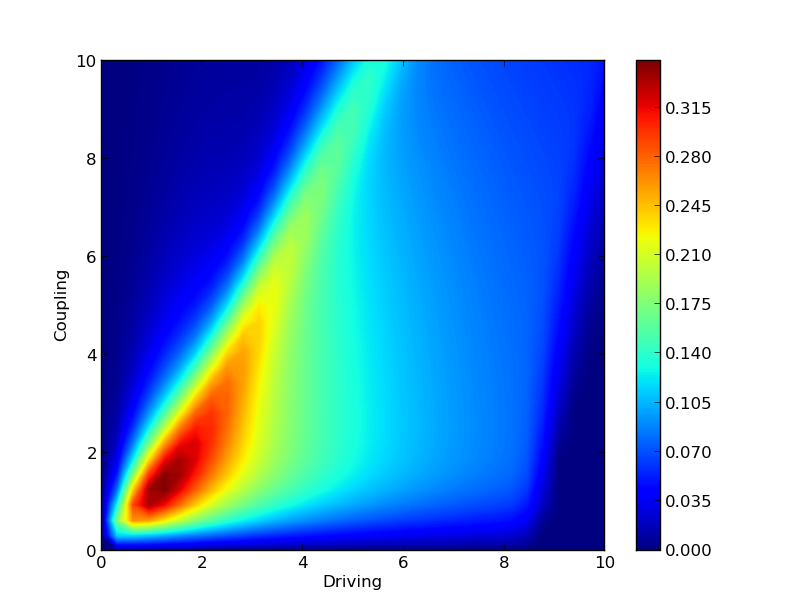}
   \end{tabular}
   \end{center}
   \caption[example]
%>>>> use \label inside caption to get Fig. number with \ref{}
%   { \label{fig:example}
{Log negativity as a function of driving field and atom-field coupling for $\kappa = 1$ and $\gamma=0.1$.}
   \end{figure}
   
Note that we have not set $\gamma=0$ as in the work of Alsing et. al. \cite{Alsing-92-DSS,Alsing-91-DressedState}. The phase transition seems to be present as long as $\gamma$ is the smallest rate in the problem.
 
\section{Conclusion}
We have investigated entanglement in an open quantum system by calculating the
log negativity of a system treated via quantum trajectory theory. We believe
that this methodology is useful in describing the behavior of entanglement in
open systems. We stress that a different unraveling of the master equation
via different choices for $\mathcal{S}$ will result in different results. We
have found that there is an optimum value of the driving field for generating
entanglement. This value is at or below the value at which the atomic
inversion begins to saturate. At higher field values, the atom begins to
uncouple from the field mode. This decoupling is due to spontaneous emission,
which is maximized when the atom is saturated. We have shown that a steady
state entanglement is well defined in the ensemble average over trajectories,
and that the entangement typically exhibits a larger value transiently than it
does in steady state. Additionally, we find that for large driving fields
there are detunings for which multi-photon resonances occur. Here, we find
behavior similar to that of a two level system. As the driving field is raised
to the value $2 \epsilon = g$, there is a phase transition to a bimodal
regime. We find that near the phase transition atom-field entanglement is
appreciable, and is maintained into the bimodal regime where its maximal value
is obtained.

The authors would like to thank H. J. Carmichael and J. P. Clemens for extremely useful discussions.

\bibliographystyle{apsrev4-1}
\bibliography{NSF15}

\end{document}